\documentstyle[twoside,fleqn,espcrc2]{article}

\def\MSB{{\rm \overline{MS\kern-0.14em}\kern0.14em}}

\newcommand{\AmS}{{\protect\the\textfont2
  A\kern-.1667em\lower.5ex\hbox{M}\kern-.125emS}}

\title{The QCD Spectrum }

\author{Chris Michael\address{D. A. M. T. P.,
        University of Liverpool, \\
        Liverpool L69 3BX, U.K.}
     \thanks{To be published in the Proceedings of Lattice 1994,
Nucl. Phys B (Proc. Suppl.), Liverpool Preprint LTH 340,
heplat/9412032, December 5, 1994}
        }

\begin{document}

\begin{abstract}

Recent results in light hadron spectroscopy are reviewed. Attention
is given to the requirements of precision determinations in
lattice gauge theory.  Different methods for extracting the
running coupling constant $\alpha_S$ are compared.
Some recent results in determining the nature of the interaction
between static quarks are also discussed.

\end{abstract}

\maketitle

\section{INTRODUCTION}

In principle lattice Monte Carlo methods are capable of precision
evaluations of the QCD spectrum. Furthermore one can study the  effect
of varying the fundamental parameters, namely the quark  masses, from
their experimental values.

In order to organise the  discussion, I present the current state of QCD
spectrum calculations  by comparing them to the ideal target of perfect
accuracy. The topics  to be covered are the control of the
discretisation errors, finite volume errors,  quark mass errors and
ground state extraction errors.

In a later  section I discuss the link to perturbation theory.  This  is
relevant for an accurate extraction of the running coupling of QCD. It
is also relevant for other perturbative contributions, such as  the
matching coefficients ($Z$) needed to relate lattice matrix elements to
continuum matrix elements.

The interaction between static colour sources is discussed since
recent results shed light on the confinement mechanism.

\section{PRECISION QCD ON A LATTICE}

One aim of lattice QCD is to be able to reproduce accurately the
hadron spectrum. This will then give confidence in the method and
will allow evaluations of new quantities of interest such as
matrix elements. A study of the effect of varying the quark
masses is also of great interest.  Because of the extreme
computational resource needed for full QCD simulation, progress
can be made by studying quenched and partially quenched
systems as well.

There have been new dynamical fermion calculations
reported~\cite{columbia,apefull} this year. A growing trend is to
compromise between  quenched and dynamical calculations by using partial
quenching. This  implies using valence quarks with different properties
(mass, or  discretisation) from the dynamical (sea) quarks present in
the  vacuum.  The quenched approximation itself is still a rich subject
of study and much remains to be done.

The ultimate aim of lattice QCD calculations is to extract physical
results  applicable to the continuum limit. To organise the presentation
of the current state of the subject, I consider in turn the  various
approximations inherent in the lattice approach.

\subsection{Discretisation Errors}

The lattice spacing $a$ is introduced to regulate the field theory  of
QCD. Then dimensionless ratios of mass values determined from lattice
calculations  will have errors of order $Ma$ or $\Lambda a$, in
general, where  $M$ is a mass value and $\Lambda$ represents a typical
QCD momentum scale. The continuum limit is obtained by extrapolating
such ratios to $a=0$. Even if only states with light quarks (u,d and s)
are studied, the  masses can be large: the $\Omega^-$ has a mass of 1.67
GeV. So, a priori, corrections of order $Ma$ can be large for current
values of the lattice spacing of $ a^{-1} \le 3 $ GeV.

Note that dimensionless ratios such as $M/\Lambda$ will have
discretisation errors of order $g^2$ or $\log(a)$ since $\Lambda$ is
defined  through matching with perturbation theory. Thus one should not
use  such ratios to obtain a continuum limit.

\begin{figure}[t]
\vspace{2.4in}
\includegraphics{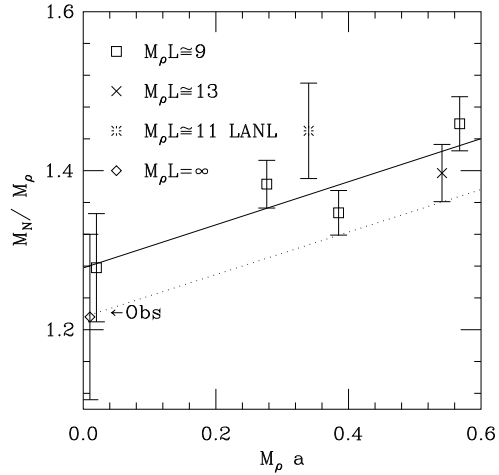}
\caption{\label{nrho} The nucleon to $\rho$  mass ratio from quenched
results~{\protect\cite{gf11,LANL}} extrapolated to the chiral limit.}
\end{figure}

For the study of hadrons made of quarks, the Wilson discretisation  of
fermions is simple and flexible. It has errors of ${\cal O} (a)$. As an
illustration, consider the nucleon to $\rho$ mass ratio. The most
extensive results have been obtained in the quenched approximation.
Then keeping the lattice volume constant in physical units (at
$M_{\rho} L \approx 9$), and  after  extrapolating the quark masses to
the physical light quark masses, the ratio has been fitted~\cite{gf11}
as
 \begin{equation}
 {M_N \over M_{\rho}} = 1.28(7)+0.27 M_{\rho} a
\end{equation}
 \noindent as illustrated in fig~\ref{nrho}. This rate of $a$-dependence
implies that  an extrapolation in $a$ is needed even with data at $\beta
= 6.17$ on  $32^2 \times 30 \times 40$ lattices where $M_{\rho}a \approx
0.3$. The results shown in  fig.~1 from the GF11 group~\cite{gf11} are
the most comprehensive available. Nevertheless, the statistical errors
are quite large so that,  even  in the quenched approximation, results
for $M_N / M_{\rho}$ are  not yet precise.  Thus although the GF11 data
are consistent after extrapolation in $a$ and $L$ with the  experimental
ratio of 1.22, much work still needs to be done to  get an accurate
calibration of the quenched approximation value. A recent analysis by
the  LANL group~\cite{LANL} using $32^3 \times 64$ lattices  at
$\beta=6.0$ has also produced relevant data and this  is shown in fig~1.
Even with 80 configurations, they are unable to reduce their errors
sufficiently to provide a stringent cross check.


The situation for staggered fermion discretisation in the quenched
approximation is summarised in  the review by Gottlieb~\cite{stag}.
Again current data are consistent with an $a$-dependence which yields
a continuum limit close to the experimental ratio of 1.22.

Given the substantial ${\cal O} (a)$ corrections observed with Wilson
fermions, much effort has gone into formulations that improve the
discretisation errors. The SW-Clover fermionic action has ${\cal
O}(\alpha_S a )$ errors with a correct treatment of  observables
{}~\cite{clover}.  Simulations using this action~\cite{ukqcd,ape}  at
$\beta=6.0$ and $6.2$ are  not yet of sufficiently high statistics to
enable  $M_N / M_{\rho}$ to  be determined accurately. In fact I  return
to discuss the current state of the UKQCD nucleon data when I  discuss
ground state extraction later.

For heavy quarks, the hadron spectrum is even heavier and ${\cal O}(Ma)$
corrections are even more important. For c quarks, the SW-clover
formulation is promising and has been successfully used.  For b quarks,
it is feasible to use a formalism which treats the heavy quark
explicitly: non-relativistically (NRQCD), as static quarks, or by a
discretisation of  heavy quark effective theory. Such heavy  quark
calculations are covered in  talks by Sommer~\cite{sommer},
Sloan~\cite{sloan}  etc.

Another approach to improving the discretisation errors, is the
mean-field improved formalism~\cite{lm}. This provides
an impressive description of many features. What is needed is a
method to estimate reliably the residual error.

The goal of removing completely the discretisation errors is
achieved by  the perfect action. For the gluonic
sector of the theory, the standard Wilson plaquette discretisation
already has errors only ${\cal O }(a^2)$. Current approximations
to the perfect action are reported by Hasenfratz~\cite{ah}. As
well as a perfect action, perfect observables will be needed to
remove lattice artifacts such as the lack of rotation invariance
in the static potential for example.

\subsection{Finite Volume Errors}

Corrections in a spatial volume  $L^3$ can arise because of exchange  of
the lightest hadron (the $\pi$ meson of mass $m_\pi$) around the
boundary.  Finite size corrections can also occur if the radius of a
hadronic state $r_h$ is not small compared  to $L$. To remove these
corrections,  one needs
 \begin{equation}
  m_{\pi}L >> 1   \hbox{, and} \ \ r_h/L << 1
\end{equation}
\noindent The finite value of $L$ also makes momentum discrete and
for some studies it is appropriate to require the minimum
momentum $q_{min}=2 \pi /L << m_{\pi}$.

There can be surprises: in pure gauge studies the finite size effects
on the glueball spectrum are significant out to $m_G L \approx 8$ where
$m_G$ is the lightest state (the $0^{++}$ glueball). So ``$>>$'' really
means ``$>>$'' not just ``$>$'' in this case. This is understood and
arises since the dominant  finite size effect comes from flux loops
encircling the spatial  boundary (called spatial Polyakov loops or
torelons).

Moreover, these effects were predicted~\cite{capri} to be approximately
twice as  strong in full dynamical quark simulations since one torelon
(rather than two) can then mix with the  glueballs.

\begin{figure}[t]
\vspace{1.2in}
\includegraphics{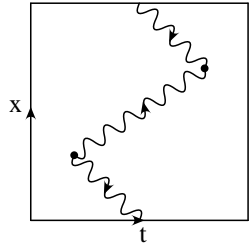}

\caption{ \label{fsef} A finite size effect}

\end{figure}

At the Lattice 1993 conference, results were presented~\cite{fse}
showing that  finite size effects on the hadron spectrum indeed appear
worse for full QCD compared to the quenched  approximation. The origin
of the effect can be seen from the example  of a meson correlation
illustrated  in fig~\ref{fsef} where one quark propagator encircles  the
spatial boundary. So effectively, the quark feels the (non-abelian)
phase of the spatial Polyakov loop. Now in the quenched approximation,
the $Z_3$ symmetry ensures that such loops have zero expectation value.
In contrast, the quark mass term in the dynamical quark case breaks
this $Z_3$ symmetry. This then allows the Polyakov loops to become
non-zero on average and so enhances their contribution to finite size
effects.

A new study of this topic was reported by the APE
collaboration~\cite{apefull}. They use anti-periodic spatial boundary
conditions on the Wilson dynamical (sea)  quarks. This `polarises' the
Polyakov loops so reducing statistical errors. They then explore the
magnitude of the residual finite size effects  by measuring $\pi$ and
$\rho$ meson masses using both periodic and anti-periodic  spatial
boundary conditions for the valence quarks. With $L \approx  1.5\,$fm,
they find finite size effects on the masses of only a few per cent.

\subsection{Quark mass extrapolation}

The requirement is to incorporate fully quarks of physical mass in  the
QCD vacuum. Because of the large computational resource needed,  typical
studies are limited to spatial sizes of $16^3$ and  quark masses
corresponding to $m_{\pi}/m_{\rho} \approx 0.3$. For technical
reasons, one of the preferred methods is to  include two flavours of
degenerate light quarks using the staggered fermion  discretisation.

In most simulations, the main effect of including dynamical quarks  is a
shift of the $\beta$ value. Thus, instead of comparing at the same  bare
$\beta$ value, one should compare at the same lattice spacing (ie  same
$ma$ value). Then  ratios of masses  appear very little changed compared
to the quenched case. Since the present quenched spectrum is consistent
with the experimental  data, it is to be expected that adding dynamical
quarks will  not change this state of affairs.  One expected effect of
dynamical quarks  is to modify the small distance forces. This region is
amenable to  perturbative evaluation and, because of the  factor
$33-2N_f$ in the denominator of the running coupling, a stronger
interaction is expected at short distance.  This can be tested by
looking at the force between static quarks at short distance~\cite
{bitar}.

An extrapolation of the light quark masses to the physical value  is
needed. For full QCD, data are not sufficiently precise to  explore this
extrapolation accurately. Further work is needed  to check whether the
extrapolation is smooth: some ``threshold effect'' might occur as the
quark mass is reduced (the opening up of the  decay $\rho \to \pi+\pi$
has been suggested).

For studies of the quenched (and partially-quenched where the  valence
quarks are different in mass from the dynamical or sea quarks)
approximation, the limit of small quark masses is potentially  tricky.
This has been studied in generalised chiral effective  theories and is
reported on by Gupta~\cite{gupta}.  To  circumvent these problems
associated with extrapolation to the chiral limit, one  way is to
present lattice results at  the quark mass used. This  is appropriate
for comparing between lattice calculations (eg. the  Edinburgh plot) but
may also be used to compare with experiment  in principle. From lattice
results for hadron masses for a range of quark masses, it should be
possible to develop an accurate semi-empirical mass formula  relating
the hadron mass to the masses of its quark ingredients. Then
experimental data  for the combinations of quark masses (u, d and s)
accessible can be used to fix the coefficients in such a  model. Thus
lattice data would be used to validate a quark-model mass formula
which could then be used for direct comparison with lattice results.

While computational resources limit full QCD work, there is still
a lot of physics that can be studied in the quenched approximation.
The quenched approximation has spontaneous chiral symmetry breaking
and gives a surprisingly good overall description of the QCD
spectrum.  It is even possible to estimate matrix elements
within the quenched approximation which enable one to study
processes beyond that approximation. Examples are
the decay $\rho \to \pi+\pi$; glueball decay and the $\eta'$ mass.

Here I discuss some of these analyses to illustrate the possibilities.

\subsubsection{Glueball decay}

\begin{figure}[t]
\vspace{1.0in}
\includegraphics{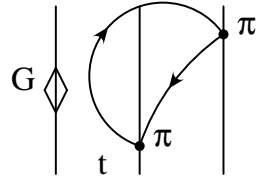}

\caption{ \label{gbf} The glueball $\ \pi \, \pi$ correlation}

\end{figure}

The glueball spectrum has been studied in the quenched
approximation~\cite{ukqcdglue,gf11glue}
and the lightest glueball is found to be the $J^{PC}=0^{++}$ state
at around 1600 MeV with the next heaviest state of $2^{++}$.

Within the quenched approximation, the lightest glueballs
cannot decay.  In full QCD, they can decay to mesons: for
instance a pair of pseudoscalar mesons. There will also be
mixing between glueballs and mesons with quark content.

There have been 20 years of unsuccessful experimental  attempts to
identify a glueball. To guide this search, it is important to estimate
the magnitude of the matrix elements between glueballs and  light  quark
mesons. Then, should the matrix elements be small, the assignment  of
glueball status could be made with confidence. Candidates  for the
$0^{++}$ glueball are the $f_0(1590)$ meson with  width $180\pm17$ MeV
and the $f_J(1710)$ meson (was called $\Theta(1690)$)  which has a
decay width of $140 \pm 12$ MeV.

{}From the point of view of lattice evaluation, such matrix elements  are
hard to measure because of a mis-match between the  techniques used to
extract glueballs and those used for light quark mesons.  A clear
glueball signal can only be  obtained by analysing many configurations
(typically thousands  of independent configurations with all time-slices
used as  a source).  Because of the computational overhead of inverting
the  fermionic matrix,  quark propagators are usually only evaluated
for hundreds of configurations and from only one source per
configuration. Furthermore the glueball signal only extends for a few
time steps  whereas a pseudoscalar meson can be followed out to very
large time.

\begin{figure}[t]
\vspace{2.2in}
\includegraphics{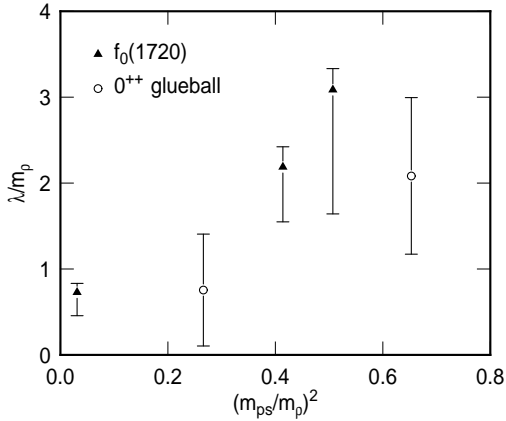}

\caption{ \label{decayf} The glueball decay coupling.}

\end{figure}

A pioneering attempt to tackle this problem has been made by  the GF11
group~\cite{gf11decay}. They use lattices of spatial size $16^3$  at
$\beta=5.7$. Using Wilson quarks with $K=0.165$, they measure  3800
configurations. They are able to isolate the $0^{++}$ glueball,  pion
and two pion states. The two pion state is calibrated by  exploring
expectation values of 4 pion operators. Then the  matrix element between
a glueball and such a two pion state is  studied - see fig~\ref{gbf}. In
principle the intermediate state has  many different contributions:
glueball, excited glueballs,  $\pi(k)+\pi(-k)$ with relative momentum
$k=2\pi n/L$ and  $0^{++}$ mesons. To separate these contributions by
studying  the $t$-dependence of the observable is difficult in principle
and hopeless in practice. Since a reasonable signal  only survives out
to $t=1$ with their current statistics, the GF11 group can make, at
best,  a very  rough estimate of the size of the matrix element.  Even
such a rough estimate can be useful: for example by indicating that a
large  glueball width is not allowed.  In the quenched approximation the
glueball does not decay and, moreover, the analysis  is performed with a
choice of quark mass such that $m(0^{++})  \approx 2m_{\pi}$.  Thus the
on-shell glueball  decay matrix element can be estimated from the
lattice measurements.

The result shown in fig~\ref{decayf} is consistent, within large errors,
with that  from the observed decay of the $f_J(1710)$ meson. Some
evidence  is also obtained that the decay matrix element increases with
increasing pseudoscalar mass - as observed experimentally.  See
ref~\cite{gf11decay} for more details.

\subsubsection{The $\eta'$ mass}

\begin{figure}[t]
\vspace{1.0in}
\includegraphics{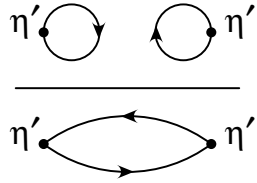}

\caption{ \label{etaf} The ratio of disconnected to connected diagrams.}

\end{figure}

In the quenched approximation, there are no quark loops in the  vacuum.
Then, since the vacuum has isospin zero,  the flavour octet and singlet
pseudoscalar  mesons ($\eta_8$ and $\eta_1$) will be degenerate if  all
quark masses are the same.  Taking the strange quark mass into account
from the $\pi$ and K   leads to  an $\eta_8$ at 567 MeV and an  $\eta_1$
at 412 MeV.  Experimentally the mixing is such that approximately the
$\eta_8$ is  the $\eta(547)$, while the  $\eta_1$ is  the heavier
$\eta'(958)$ meson. Thus in the quenched approximation,  the
contribution from the $\eta'$ will have a mass which is much too low.
This, incidentally, is at the root of  the modified chiral behaviour in
the quenched approximation.

Within the quenched approximation one may study the mixing
between quark loops and the $\eta$ mesons explicitly. Let
$P_8=(p^2+m_8^2)^{-1}$ be the quenched $\eta$ propagator (which
is the same as the $\eta'$ propagator in the quenched approximation),
Then  consider the effect of a quark loop $Q$ on the $\eta'$ propagator
$P_1=(p^2+m_1^2)^{-1}$:
\begin{eqnarray}
P_1 &=& P_8 - P_8 Q P_8 + P_8 Q P_8 Q P_8 + \dots \nonumber \\
    &=& {1 \over P_8^{-1} + Q } = {1 \over p^2+m_8^2+m_0^2}
\end{eqnarray}
 \noindent Thus $m_1^2=m_8^2+m_0^2$ where $m_0^2$ is the matrix element
of a  quark loop $Q$ in the quenched vacuum. This  mixing  works to
increase the mass of the singlet -  so giving a chance to obtain the
experimental $\eta' - \eta$  mass splitting.  Lattice methods to explore
this  have been proposed~\cite{eta}. The basic idea is to compare  the
disconnected $\eta'(0) \eta'(t)$ correlator with the  connected
correlator - see fig~\ref{etaf}. The ratio is proportional to  the
required mixing matrix element $m_0^2$. The technical  difficulty is to
measure the disconnected correlator: an ingenious  method was used to
cut down the number of propagator inversions by putting sources at each
spatial site on a timeslice and allowing  gauge invariance to select the
required contribution.  An analysis using  Wilson valence quarks at
$\beta=5.7$ on  $12^3 20$  and $16^3 20$ lattices yields mixing $m_0=751
\pm 30$ MeV to be compared to the experimental value
 \begin{equation}
     m_0^2 = m_{\eta'}^2 - m_8^2 \approx (852\, \hbox{MeV})^2
\end{equation}
\noindent So indeed quenched QCD does yield a reasonable understanding
of the $\eta'$ mass.

\begin{figure}[t]
\vspace{2.4in}
\includegraphics{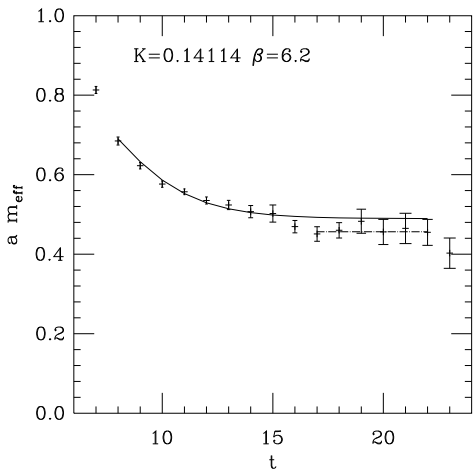}

\caption{ \label{n12} The nucleon effective mass versus $t$ from UKQCD
LL data with one and two exponential fits.}

\end{figure}

The Witten-Veneziano formula relates the $\eta$ mass mixing
to the topological susceptibility $\chi$ by
\begin{equation}
m^2_0 = { 2 N_f \, \chi \over f^2_{\pi}}
\end{equation}
 \noindent Measurements of the topological susceptibility in lattice
gauge  theory are plagued with uncertainties from lattice artifacts.
Using the cooling method on the same lattices as above, a  result of
$m_0^2=(1146 \pm 67 \,\hbox{MeV})^2$ was obtained~\cite{eta}. This is
qualitatively similar to the value obtained by studying  the mixing
directly and is further confirmation that the  underlying physics of the
$\eta'$ is understood.

\subsection{Ground state extraction}

\begin{figure}[t]
\vspace{2.4in}
\includegraphics{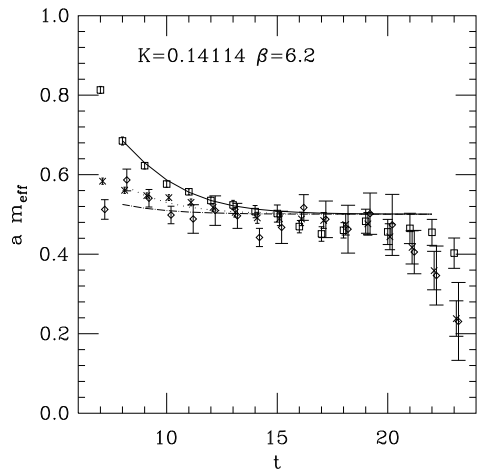}

\caption{ \label{n3} The nucleon effective mass versus $t$ from UKQCD
preliminary data for LL(squares), SL(crosses) and SS(diamonds) with
a two exponential fit. }

\end{figure}

The euclidean time formulation implies that eigenstates of the
transfer matrix contribute to hadronic correlators
$<\!H^a(0)H^b(t)\!>$ as
\begin{eqnarray}
 C_{ab}(t) & = & <\!H^a(0)H^b(t)\!>  \nonumber \\
& = &  h_0^a h_0^b (e^{-m_0t}+e^{-m_0(L-t)}) + \nonumber \\
 & &   h_1^a h_1^b (e^{-m_1t}+e^{-m_1(L-t)}) + \dots \end{eqnarray}
 \noindent where $h_i^c$ is  the amplitude to produce eigenstate $i$
from operator $H^c$.  The expression is written assuming periodic
boundary conditions in time for bosonic eigenstates.  The data  is
usually presented by computing an effective mass from  $m_{\rm
eff}(t)=-\log(C(t)/C(t-1)$. The  ground state contribution dominates
$m_{\rm eff}(t)$ at large $t$ since  $m_0 < m_i$ for all $i \ne 0$.  In
practice, extracting the  ground state mass $m_0$ from data with
statistical errors which increase  with $t$ is subtle.  I wish to
emphasize  that a good procedure is:

(i) a multi-exponential fit to the  widest acceptable $t$-range with

(ii) several hadronic operators $H^a$ to stabilise the fit.

 One way to understand this guide is that the ground state is only
determined accurately when an estimate of the first excited state  is
available. This is necessary since the energy difference controls the
rate  of approach of $C_{ab}(t)$ to the expression given by the ground
state   component alone. However, fitting 2 (or more) exponentials to
just  one function  $C(t)$ is not very stable: better is to have several
 such functions (provided that they do indeed have different  relative
amounts of ground state and excited state).

As an example of this discussion, I use some preliminary UKQCD data  for
the nucleon mass determination. With just one quantity  measured (the
local source - local sink: LL), a plateau in  the effective mass was
found as shown in fig~\ref{n12}.  This shows a  typical problem: the
effective mass decreases out to large $t$-values. I have called this the
``droop''. It is difficult to be sure  whether such a behaviour is
statistically significant or not.  For instance, a 2-exponential  fit to
the same data gives a significantly higher value of $m_0$.  A resolution
of this dichotomy comes from adding more  measurements: SL and SS where
S refers to a smeared source or sink. The combined fit to the 3 types of
data is shown in fig~\ref{n3} and strongly  favours the 2-exponential
fit with the higher $m_0$ value.

\subsubsection{Correlations among data}

It is widely known that, particularly for hadronic measurements,
the data $C_{ab}(t)$ are strongly correlated statistically. For
example $C_{ab}(t)$ and $C_{ab}(t+1)$ are often 99\% correlated.
By this I mean that each fluctuates among lattice samples within
its error, but that when the value at $t$ is above average
then so will be the value at $t+1$. This means that each $t$ value
is very far from statistically independent. The standard way to
accommodate this is to use `correlated $\chi^2$'.

The drawback of this method is that it can become unstable if
the number of lattice samples $N$ is insufficient when there are
$D$ types of data being fitted (ie $D$ is number of $t$-values
times number of types of measurement).  The expected $\chi^2$
per degree of freedom
from such a correlated $\chi^2$ fit is~\cite{cor}
\begin{equation}
{ \chi^2  \over \hbox{d.o.f.}}=1 + { D+1 \over N} + {\cal O}(N^{-2})
\end{equation}
 \noindent Moreover, if $D \ge N$, then $\chi^2$ becomes infinite since
the  correlation matrix among the data is non-invertible.

This is a problem in practice: if 15 $t$-values are to be fitted  with 3
types of observable then $D=45$ while $N$ is typically  between 50 and
200. The recommended way forward is to parametrise  the correlations
among the data with fewer parameters than the  general case (which  has
$D(D-1)/2$). This should ensure a stable evaluation of the inverse of
the correlation matrix and, in particular, avoid the problem of
spurious small eigenvalues of the sample correlation matrix.   Various
suggestions have been made  to accomplish this:  an SVD
inverse~\cite{SVD}, 3 and 5-diagonal inverses~\cite{cmamck},
averaging of small eigenvalues of the correlation matrix~\cite{cmamck},
etc. The methods used will need to take into account the nature of the
correlations among the data in each application and then model them in a
stable  way~\cite{cmamck}. A program incorporating a broad selection of
these  methods is available from the author.

The alternative to such modelling of correlations is to have enough
independent data samples $N$. Data sets containing of the order of
thousands of configurations analysed are going to be needed for
precision hadron spectroscopy on a lattice.

\subsubsection{Efficient hadronic operators}

The strategy for fitting the data to extract mass values involves
using several hadronic operators. For example local and smeared
operators were used in the previous illustration. The availability
of several operators allows a variationally improved operator to
be constructed for the ground state. It is, never the less, necessary
to select such an operator basis wisely.

\begin{figure}[t]
\vspace{2.5in}
\includegraphics{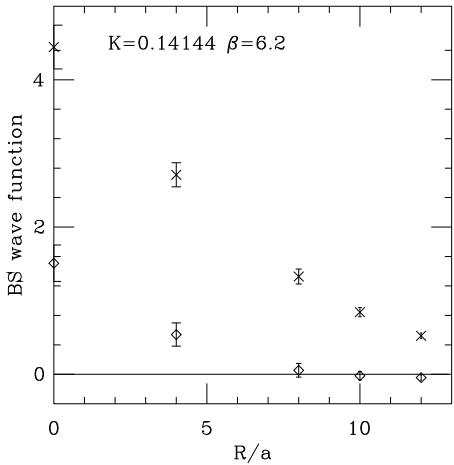}

\caption{ \label{wf} The Bethe Saltpeter wave function for
the $\rho$ and $\rho'$ from ref~{\protect\cite{lacock}} }

\end{figure}

It is important to explore this choice of operators to find  efficient
ones. The definition of efficient in this context  is that the
contribution of excited states is minimised relative  to the ground
state contribution.  Non-local operators modelled on  heavy quark bound
states are promising candidates: they would  be the optimum choice in
the limit of heavy quarks and so may  still be  appropriate for lighter
quarks. The construction used  is to join the quark and anti-quark in a
meson by a straight  colour flux path of length $R$. Following
experience of  potential studies~\cite{smear}, this flux path is made
from smeared links. The resulting overlap amplitude
 $<\!0|q(0)U(0 \to \! R) \Gamma_M \bar{q}(R)|M\!>$
 is often referred to as the Bethe-Saltpeter amplitude of meson M
(where $\Gamma_M$ is $\gamma_i$ for the $\rho$, etc). This has been
studied before~\cite{bswf,tn} to learn  about the spatial distribution
of hadrons. The new  observation~\cite{lacock} is that the same quantity
for the first excited state (eg. $\rho'$) has a node. Thus choosing the
separation $R$ at this node will  give an efficient operator to make the
ground state with no  contamination from that first excited state. As
illustrated in  fig~\ref{wf}, this  node is at $R \approx 3
\,\hbox{GeV}^{-1}$. Results for baryons are  also qualitatively
similar~\cite{lacock}.

\section{THE QCD RUNNING COUPLING}

The world average of the QCD running coupling compiled by
Webber~\cite{webber} is

$\alpha_{\MSB}^{(5)}(M_Z)=0.117(5).$

\noindent This average includes input from lattice analyses. Indeed
the lattice determinations have the smallest errors quoted. Since
the precise value of $\alpha_S$ is important for phenomenology and
for attempts to look for evidence of models beyond the standard
model, it is important for the lattice community to understand
fully the source of errors in the lattice determinations. This
is highlighted by the change from the value 0.108(6) quoted~\cite{aida}
at Lattice 1993 to a value of 0.115(2) quoted~\cite{nrqcd}  this year.

There are two requirements to measure $\alpha_S$ on a lattice:

(i) an accurate match to perturbation theory

(ii) an accurate lattice energy scale set by some experimental input
(eg. $m_{\rho}$, $V(R)$, 1P-1S heavy onia, etc)

\subsection{Perturbation theory on a lattice}

First I discuss the requirement that perturbation theory is
accurately reproduced on the lattice so that $\alpha_S$ can be
defined. Consider first perturbation theory in terms of the bare
lattice coupling $g^2=6/\beta$ with $\alpha=g^2/4\pi$.
The plaquette action $S_{\Box}$ has been calculated~\cite{pisa}
for $N_f=0$ to order $\alpha^3$ for the Wilson  gauge action.
For SU(3) colour,
\begin{eqnarray}
\alpha_{\Box} & \equiv & -{3 \over 4 \pi} \log(S_{\Box})
\nonumber \\
& = & \alpha(1+3.37 \alpha + 17.69 \alpha^2 +\dots)
\end{eqnarray}
 \noindent Now at $\beta=6.5$, $\alpha=0.07346$ and the last expression
evaluates to 92\% of the first expression obtained from the  measured
lattice plaquette action. The $\alpha^3$ term contributes 7\%.  So we do
not have a good convergence of perturbation theory. Increasing $\beta$
substantially to decrease $\alpha$ is not feasible with  present
computing techniques.

The way forward has been known for a long time.
A {\it physical
definition} of $\alpha$ should yield a perturbative series with
coefficients which are much smaller than in the case of  the bare
coupling. Consider several such physical definitions:

(i) $\alpha_V$ from $V(q)$ or $S_{\Box}$~\cite{lm}.

(ii) $\alpha_{q \bar{q}}$ from the force $dV/dR$~\cite{force}.

(iii) $\alpha_{SF}$ from the Schr\"odinger functional~\cite{sf}.

\noindent A variant of $\alpha_{SF}$ with
twisted boundary conditions has recently been used~\cite{twist}.

Consider the first case, where the Fourier transform of
the static potential is used as a definition:
$3 q^2 V(q)=-16 \pi \alpha_V(q)$. Then the logarithm of the plaquette
action can be re-expressed in terms of this definition:
\begin{equation}
 \alpha_{\Box} =   \alpha_V
(1-(1.19+0.07N_f) \alpha_V+ {\cal O}(\alpha^2_V) )
\label{pl-v}
\end{equation}
 \noindent where $\alpha_V$ has scale $q=3.41/a$ \cite{lm}. Indeed the
coefficient  of $\alpha^2$ is now ${\cal O}(1)$. The expectation is that
all  coefficients in such an expansion are ${\cal O}(1)$. For our
present purposes it is the coefficient of the $\alpha_V^3$ term above
that is assumed to be ${\cal O}(1)$.

This coefficient can be obtained  by evaluating the perturbative
connection between the bare and $V-$scheme coupling to order  $\alpha^3$
since the link between $S_{\Box}$ and bare coupling  is known (though
only for $N_f=0$).   Indeed, since the calculation of the connection
between  $\alpha_{\MSB}$ and the bare coupling is under way
{}~\cite{weisz}, it will soon be possible to determine $\alpha_{\MSB}$
directly from $S_{\Box}$ to relative accuracy of ${\cal O}(\alpha^3)$.
Meanwhile, the connection of the $\MSB$ and V schemes  is only known to
lower order. For example for  SU(3) of colour,
 \begin{equation}
\alpha_{\MSB}=\alpha_V (1-(0.822-0.088N_f) \alpha_V +
{\cal O}(\alpha_V^2)) \label{msbv}
\end{equation}

When comparing $\alpha$ in different schemes, the scales in the  two
schemes can in principle be chosen differently. Possible choices are  to
take them equal, choose them so that the ${\cal O}(\alpha^2)$ term  is
exactly zero, or some compromise. From a comparison of the
Schr\"odinger Functional and Twisted Polyakov methods  in SU(2), there
is a definite preference~\cite{alpha} for the prescription of  choosing
the relative scale (as the ratio of  $\Lambda$ parameters) so that the
the ${\cal O}(\alpha^2)$ term  is exactly zero. As a further example of
this ambiguity, eq~\ref{msbv} can  be expressed~\cite{nrqcd} as
 \begin{equation}
\alpha_{\MSB}(q)=\alpha_V(e^{5/6}q) (1+0.637 \alpha_V  +
{\cal O}(\alpha_V^2))
\end{equation}
 \noindent The difference between these expressions is ${\cal
O}(\alpha^3)$ and so a preference can only be made when the
$\alpha^3$ calculation is completed.

Until the $\alpha^3$ calculations are completed,  some cross checks are
worth making. One approach~\cite{dimm} is to use Monte Carlo evaluation
on lattices with a wide range of $\beta$  to estimate the size of the
$\alpha_V^3$  correction in eq.~(\ref{pl-v}) . The method is to compare
the  extraction of $\alpha_V$ from $1 \times 1$, $1 \times 2$,  $1
\times 3$ and $2 \times 2$ Wilson loops. Then, relative to the (unknown)
coefficient of the $1 \times 1$ extraction, the other cases have
$\alpha^3$ coefficients  of 0.36, 0.53 and 0.97 respectively. This is
indeed confirmation  that coefficients ${\cal O}(1)$ are plausible.

Another cross check is to use the recent perturbative evaluation~\cite
{wolff} of $\alpha_{SF}$ in terms of the bare coupling to obtain
for SU(2) of colour:
\begin{equation}
\alpha_{\Box}=\alpha_{SF}(1-0.737\alpha_{SF}+0.039\alpha_{SF}^2 + \dots)
\end{equation}
with scale $a=q^{-1}=L$. This is the first relationship  to be obtained
to this order between two physically defined lattice  couplings.  The
coefficients are indeed less than one, although with the caveat  that
SU(2) colour tends to have smaller coefficients than SU(3).

Besides evaluating $\alpha_V$ from $S_{\Box}$, it can be directly
obtained from the static quark potential~\cite{klassen}. Present results
are  compatible between the two methods within the inherent errors of
each.

Many phenomenological models have been proposed to obtain an  improved
definition of $\alpha$ in lattice studies~\cite{parisi,kp,lmmf} by
modifying the bare coupling. These suggestions were constructed with the
intention of  rendering the perturbative coefficients of ${\cal O}(1)$.
In this  respect they are similar to the $\alpha_V$ scheme outlined
above but, because  they are empirical in origin, they are less
appropriate to a precision analysis of $\alpha$.

Since $S_{\Box}$ is the shortest distance quantity defined  on a
lattice, it corresponds to  the highest  momentum. It is also easy to
measure accurately. Thus it is  a very popular starting point for
extracting $\alpha$. Indeed,  $\alpha_{\MSB}$ can be obtained from the
measured $S_{\Box}$ with  error ${\cal O}(\alpha^3)$.  In order to check
that  perturbation theory is accurate, the coefficient of this
$\alpha^3$  term should be  evaluated (this is in progress) and then the
values of  $\alpha_{\MSB}$  from different lattice spacings should be
compared with the perturbative evolution through the beta-function. If
this perturbative evolution were accurate above some scale $q=1/a$, then
one  would have confidence in this method of extracting $\alpha$.  For
the other methods of defining and extracting $\alpha$, the same checks
are needed and they have been thoroughly explored for the  Schr\"odinger
Functional scheme in the quenched approximation~\cite{sf}. This check of
the evolution is essential to control possible  non-perturbative effects
which I now discuss.

\subsection{Non-perturbative contributions}

Since the method employed is to extract $\alpha$ from lattice
measurements, there may be a non-perturbative component present  in
principle. Note that such non-perturbative components are also  equally
important in continuum/experimental determinations  of $\alpha_S$. I
shall illustrate this problem  by considering the ``force'' definition
of the coupling. For SU(3) colour with $N_f=0$, this is
 \begin{equation}
\alpha_{q \bar{q} } (q) = {3 \over 4} R^2 {dV \over dR} \ \hbox{with}
 \ q={1 \over R}
\end{equation}
 Now, since the force $dV/dR$ has a non-perturbative contribution from
the string tension $K$, one could equally well introduce a modified
definition by  subtracting it: $\tilde{\alpha}=\alpha_{q \bar{q} } -3 K
R^2/4$. Since  $\alpha \approx b_0/\log(R^{-2})$, the two definitions
of $\alpha$ differ by a term proportional to $\exp(-b_0/\alpha)$. This
is a non-perturbative difference: the two definitions agree to  all
orders of perturbation theory. For $q=R^{-1} >> 10$ GeV, this
non-perturbative piece is negligible ($ < 1 \%$). However, one should
keep in mind that different ``physical'' definitions of $\alpha$  could
have different non-perturbative components. For example  when extracting
$\alpha_V$ from $S_{\Box}$, the typical momentum  scale is large
($q^*=3.4/a$) but it arises from a cancellation  of $q_i=0$ and
$q_i=\pm\pi/a$. So non-perturbative effects could  arise even though
$q^*>10$ GeV.

\subsection{Comparison of $\alpha_S$ determinations}

I now make a comparison of several methods of extracting
$\alpha_{\MSB}$ from lattice measurements. To make the  comparison as
close as possible, the scale is set in each case from the  best measured
quantity (the static quark potential at moderate  interquark
separations) with  the definition $r_0^2 dV/dR|_{r_0} =1.65$. This
definition~\cite{sommer}  corresponds  to $r_0 \approx 0.5\,$fm and to
$\sqrt{K} r_0 \approx 1.2$. To have the most accurate lattice data, I
use the quenched determinations for SU(3) colour.

(i) Then at $q=37/r_0$ (approx 15 GeV), the Schr\"odinger functional
method obtains~\cite{sf}

$\alpha_{\MSB}=0.1108(23)(10)$

\noindent where the first error
is from the scale and the second is assuming the $\alpha^3$
correction has coefficient $\pm 1$.

(ii) The method of extracting the coupling
from the plaquette action $S_{\Box}$ (via $\alpha_V$) yields

$\alpha_{\MSB}=0.1151(5)(13)$

\noindent where data~\cite{ukqcdpot} at $\beta=6.5$ has been used  which
corresponds to $q^*=38.3/r_0$ and this has been run to  $37/r_0$ using
2-loop formulae.

(iii) The last method uses the force and,  at the highest $\beta$-values
currently available~\cite{ukqcdpot} of 6.5, corresponds  to $q < 7/r_0$
only. However, it gives results for a  range of  $q$-values consistent
with two-loop running. Assuming  two-loop running to $37/r_0$, then
yields

$\alpha_{\MSB}=0.1180(21)(13)({+0 \atop -60})$

\noindent where the last error
term comes from the possible non-perturbative effect of removing
the string tension before running the coupling. There is no sign
in the data for any such non-perturbative effect but its absence can
only be confirmed by going to smaller lattice spacing.

These values are in reasonable agreement with each other. The ${\cal
O}(\alpha^3)$  terms need to be calculated to refine the comparison
further (this will remove the 1\% error from assuming the coefficients
are within $\pm 1$).  A guide to the possible size of the ${\cal
O}(\alpha^3)$  terms comes from comparing the different definitions  of
$\alpha$ at different scales so that the $\alpha^2$ terms vanish -
rather  than at the same scale. Some evidence
exists~\cite{twist,wolff,alpha} that this prescription works well for
SU(2). For the Schr\"odinger functional method, this would then yield

$ \alpha_{\MSB}=0.1141(24)(15) $

\noindent The difference from the result quoted above is approximately
$2.5 \alpha^3$ which shows that  coefficients of that magnitude are
plausible in this case. Moreover,  this scale choice gives a value in
closer agreement with the other  methods.

After this perturbative calculation, any remaining
difference in $\alpha$ values must be ascribed to non-perturbative
effects.  Since any non-perturbative  effects are rather different in
the first method (which involves the small-volume vacuum) than in  the
latter two determinations, this will be a valuable cross check.

 This comparison  highlights that caution is needed in extracting
$\alpha$ from lattice  data. Consistency as the momentum scale is
increased is essential.

\subsection{Setting the Scale}

The requirement is that dimensionless ratios of physical
quantities are independent of the lattice spacing $a$ so that
the scale can be set unambiguously. Because of uncertainties in
corrections due to quenching and to ${\cal O}(a)$ effects,  there
is still a measure of uncertainty in setting the scale.

The most accurate lattice measurement is indeed $r_0$ from  the static
potential $V(R)$ as  discussed above. This definition, unlike that in
terms of the string  tension, is also applicable to full QCD. Indeed
this is the  method of choice for comparisons between  different lattice
 evaluations. Of course, $r_0$ is not directly known  experimentally, so
other quantities are also used.

The $\rho$ mass is often chosen although the  extrapolation of $m_q \to
0$ (or the physical light quark masses)  introduces errors. The effect
of incomplete treatment of the decay $\rho \to \pi + \pi$ is most
probably  quite small on the $\rho$ mass.

Another choice is $f_{\pi}$ but the renormalisation factor $Z_A$
then needs to be evaluated. This needs an accurate perturbative
evaluation - cross checked by non-perturbative methods - see
the review by Martinelli~\cite{marti}.

A very popular choice is the 1P-1S mass  splitting in heavy onia. This
choice is supported by two encouraging  indications: (i) empirically,
the mass splitting depends  very little on the heavy quark mass and (ii)
the relevant momenta (for  instance as indicated by the wave function)
are  expected not to be so sensitive to very light quark effects. The
drawback is that for relativistic fermions $m_c a$ is comparable to 1,
so ${\cal O} (a)$ effects are potentially large. This can be avoided  by
using an effective lagrangian appropriate to heavy quarks for the
$b\bar{b}$ system: for example a non-relativistic  treatment.  This
requires a careful calibration of the effective  lagrangian.   The NRQCD
method, see ref~\cite{sloan}, involves  an effective lagrangian which
can be refined in principle by  evaluating more terms. The inherent
errors are not straightforward  to estimate.

For the quenched approximation, the (1P-1S)/(2S-1S)  ratio for $b
\bar{b}$ comes out~\cite{nrqcd} from NRQCD as 0.71 (similar to the
static potential  value~\cite{smear}) which differs from the
experimental value of 0.78. This can induce  some uncertainty in setting
the scale.

The running of the coupling reduces the dependence on the scale
as the momentum is increased:
$ \Delta \alpha/ \alpha \approx 0.2 \, \Delta q_0 /q_0$ at $q=M_Z$.
Thus a 10\% error in the scale $q_0$ results in a relative error of
2\% on $\alpha$ at $M_Z$.

\subsection{$\alpha_S$ from dynamical quark simulations}

Until recently, attempts to evaluate $\alpha_S$ from lattices  used
quenched simulation and corrected for the effects of quenching  via a
phenomenological potential approach~\cite{aida}. This certainly captured
 one of the main corrections to quenching: namely the stronger
interquark force at small separation (by a factor of $33/(33-2N_f)$).
However, corrections at moderate and large separations were  treated
empirically. This year, several groups have used  dynamical quark
simulations directly.  These have mainly  involved 2 flavours of
staggered quarks and the coupling has been  determined by relating
$\alpha_{\MSB}$ to the plaquette action  by perturbation theory.  For
example~\cite{nrqcd}, the plaquette action gives
$\alpha_V(3.41/a)=0.1785(56)$ at $\beta=5.6$ for 2 flavours of staggered
quarks  with $ma=0.01$. Here the error  comes from assuming the
$\alpha^3$ coefficient is $\pm 1$. The scale has been  set by either
$m_{\rho}$ or the 1P-1S splitting in onia. After processing the
$N_f=0$ and $N_f=2$ results to obtain $N_f=3$, in each case the result
obtained for 3 light flavours at moderate $q$ has been run to  $q=M_Z$
in the conventional way to facilitate comparisons.

Using Wilson valence quarks, the result~\cite{aoki} obtained is

$\alpha_{\MSB}(M_Z)=0.111(5)$

\noindent where an extrapolation of $S_{\Box}$ to $m_q =0$ has been
made. The  result for $N_f=2$ has been corrected to $N_f=3$ by using
2-loop running  down to \hbox{$\approx 0.5$} GeV with 2 flavours and
then back up with three. The authors  compare the quenched and
unquenched $\alpha$ values and find that they overlap at  low $q$
values. This  is some support for their method of correcting to $N_f=3$
(and  incidentally for the method used to go directly from $N_f=0$ to
$N_f=3$~\cite{aida}). A similar approach using staggered valence
quarks~\cite{ksa} shows sizable flavour breaking effects, particularly
for heavier quarks, and thus relatively large systematic errors.

Another group~\cite{nrqcd} has used NRQCD to evaluate the $b \bar{b}$
spectrum with dynamical quark configurations. They extrapolate
$\alpha^{-1}$ in $N_f$ from $N_f=0$ and 2 to get to $N_f=3$. Using a
fixed  dynamical quark mass $am_q=0.01$ at $\beta=5.6$, gives

$\alpha_{\MSB}(M_Z)=0.115(2)$

A preliminary result has been obtained~\cite{wingate} using the
same configurations but with Wilson valence fermions. The
main difference comes from the lattice scale as $a^{-1}=1.9\,$GeV
rather than 2.4 GeV as found from NRQCD. Their result is

$\alpha_{\MSB}(M_Z)=0.108(6)$

Note also that using $r_0$ to set the scale on the same
configurations~\cite{bitar} yields $r_0/a=5.2(3)$ which gives $a^{-1}
\approx 2.0\,$GeV. This agrees with the Wilson valence result  but
differs from the NRQCD value from the same configurations. The  NRQCD
value relies on perturbative evaluation of coefficients  in the
lagrangian, so non-perturbative corrections introduce an error which is
hard to  quantify. The direct full QCD determinations have scale errors
of ${\cal O}(a)$ which can be reduced by increasing $\beta$.

In view of the uncertainties of order $4\%$ found in the quenched case
with much smaller lattice spacings, it seems prudent to include  errors
at least of this size to allow for  possible non-perturbative effects
and  possible effects from the  unknown perturbative coefficients at
${\cal O}(\alpha^3)$.  Further, the scale  error is presently some
$20\%$ which implies a 4\% error in  $\alpha(M_Z)$. In summary, I
propose that at least a $6\%$ error should be ascribed to  the lattice
results. Even so, a reasonable average of current lattice determinations
would yield 0.112(7). This is still of significance since the world
average quoted above was 0.117(5).

To improve on this, calculations at increased $\beta$ are needed to
explore  non-perturbative effects and ${\cal O}(a)$ effects.  Further
variation of  $m_q$ and $N_f$  should be explored. The ${\cal
O}(\alpha^3)$ terms in the perturbative  matching also need to be
evaluated.  Several of these steps are under way and it should soon be
possible to evaluate $\alpha_{\MSB}$ at $q \approx 10$ GeV  to high
accuracy  from  lattice studies. Assuming the conventional 3-loop
evolution from  10 GeV, this  will yield $\alpha(M_Z)$ to 1\%.

\section{THE INTERACTION BETWEEN COLOUR SOURCES}

As well as  explicit study of  the spectrum and coupling of hadrons,  it
is instructive to look at qualitative features of QCD. The  confining
mechanism is the clearest example. One consequence of  confinement is
that the potential energy $V(R)$ between static colour sources  in the
fundamental representation grows with increasing separation $R$. In  the
quenched approximation, this potential continues to grow  approximately
linearly for large $R$ - as $KR$ where $K$  is the string tension.   The
name string tension comes from the observation that a string  theory can
yield such a behaviour.  To study this confining force in more detail,
there are  several topics to pursue:  a detailed measurement of $V(R)$
at  large $R$ is of interest; the spatial distribution of the energy in
the colour flux can be determined;  the spin dependence of the  long
range force can be explored.  These topics have been of interest  for
many years. Increased computational resources have  enabled major
increases in precision recently. Because it is qualitative  features
that are to be explored, it is  sufficient to use SU(2)  colour for
initial investigations. This allows considerable increase  in precision
since it is computationally faster.

As well as studying the strong force between two heavy quarks, it is
of interest to explore the much weaker forces between hadrons. These
can be studied by looking at 4 quark potentials in SU(2).

\subsection{The self-energy of the hadronic string}

 The self-energy of closed string-like flux  tubes of length $L$ (called
torelons) has been studied in SU(2) gauge theory~\cite{torel}. For  $L >
1$ fm, the  energy of such torelons was found to  have a component
$f\pi/3L$ with $f=1.00(3)$, exactly the  coefficient expected from
string fluctuations.  This is a test of the zero point fluctuations of a
string. The  string excitations themselves give rise to excited states
of the potential between static  sources and  they have been found
previously~\cite{capri} to agree well.

\subsection{The colour flux tube}

\begin{figure}[t]
\vspace{2.5in}
\includegraphics{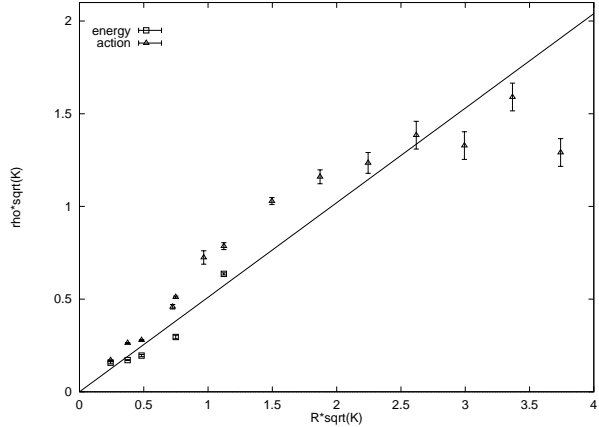}

\caption{ \label{fluxf} The half width $\rho$ of the energy
and action flux tube at source separation $R$ in units of the
string tension $K$.
}

\end{figure}

 The Wuppertal group~\cite{flux} have explored the spatial  distribution
of the colour fields between static quarks in SU(2)  with separations
out to $R \approx 2 $ fm. At smaller $R$, their results are  similar to
features seen previously~\cite{hay}.  The question of interest is the
transverse profile of the colour flux tube.  They find that the width of
this transverse flux profile  increases quite fast with $R$ until $R
\approx 1$ fm when the increase  levels off. By looking at the
differences in the distribution  of colour flux as $R$ is increased,
they find  consistency with the picture that a string-like
configuration exists for $R \ge 0.75$ fm. The data are shown in
fig~\ref{fluxf} and are consistent  with the expected increase of the
width as $\log R$ in this  region, but the  evidence is not compelling.
The breakdown into colour  magnetic and colour electric contributions is
measured and this is a  constraint on phenomenological models of the
confining flux tube.

\subsection{The spin dependent force}

    From earlier work, we know that the force between static quarks  has
a short ranged component roughly like one gluon exchange and a  long
range component which is dominantly in the $V_1'$ rather  than $V_2'$
spin-orbit term (this  corresponds to Lorenz `scalar' confinement in the
usual description). In order to get more precision, it is necessary  to
go to smaller lattice spacing to study the spin dependent potentials
since they vary very rapidly at small $R$.

 Recent work~\cite{spin} confirms the earlier qualitative results and
shows some evidence for small but significant departures from the
simple picture of one gluon exchange together with a string component.
The region studied is $R \le 0.3$ fm. In this $R$ region, the new
results confirm that the confining component is dominantly in $V_1'$. As
well as this string tension term $K$ in $V_1'(R)$, there  is also
evidence for an additional attractive $1/R^2$ piece. Since 1 gluon
exchange does not contribute to $V_1'$, such a component  could arise
from the string fluctuation term (-$\pi/12R$ in $V_0(R)$) or  from
multiple gluon exchange.

 A careful theoretical study of the  nature of the  spin-orbit force to
be associated with the long range string  fluctuation component would be
very useful since some relevant lattice data  now exists.

\subsection{The force between hadrons}

The colour force between static fundamental-representation quarks is
far stronger than the residual `strong' force between hadrons. It  has
always been difficult to explore this `strong' force in lattice
simulation because the binding energy is so much smaller than that  from
the colour force. For example for SU(2) of colour and for  2 quarks and
2 antiquarks, the difference $V_4-2V_2$ between the 4-body  potential
and that of the two pairs taken separately is typically  much less than
1\% of $2V_2$.  Nevertheless, results have been obtained  for the
binding energy for a number of different geometrical arrangements  of
the 4 static sources~\cite{v4}.  The value of the observed binding
energy can be understood  qualitatively from mixing between different
pairings of the two body potentials. One strong conclusion from this
study~\cite{v4} is that  the four body system is {\it not} well
approximated as a sum of  two body components.

\section{CONCLUSIONS}

Often the progress in a field is hard to see year by year. New
ideas become accepted gradually and less accurate methods wither
gently.  The progress is often that problems are no longer discussed,
since new methods have circumvented the obstacle.

One area of progress has been in understanding the best  way to study
the QCD coupling on a lattice. The agreement  between different methods
gives confidence. Present estimates  correspond to a value
$\alpha_{\MSB}(M_Z)=0.112(7)$ and a  clear route exists   to reduce this
error to about 1\%.

Quenched QCD is currently in good shape. The sources of error are
understood and steps are being taken to reduce them. Accurate  spectral
results in the quenched approximation are of importance,  both because
they enable us to understand QCD better and  because they provide a
calibration for calculations of matrix  elements, etc.  The full QCD
program is less advanced because of  computational limitations. Studies
so far have used rather small  lattices, rather coarse lattice spacing,
and rather heavy quarks. With future increases in computer power and
future improvements  in algorithms, it will be possible to explore  full
QCD as  thoroughly as the quenched approximation is currently studied.

\end{document}